# Role of Fabrication Errors and Refractive Index on Multilevel Diffractive Lens Performance

Sourangsu Banerji[1], Jacqueline Cooke[1] and Berardi Sensale-Rodriguez[1, *]

[1] Department of Electrical and Computer Engineering, The University of Utah, Salt Lake City, UT 84112, USA

* E-mail: berardi.sensale@utah.edu

**Abstract**

Multilevel diffractive lenses (MDLs) have emerged as an alternative to both conventional diffractive optical elements (DOEs) and metalenses for applications ranging from imaging to holographic and immersive displays. Recent work has shown that by harnessing structural parametric optimization of DOEs, one can design MDLs to enable multiple functionalities like achromaticity, depth of focus, wide-angle imaging, etc. with great ease in fabrication. Therefore, it becomes critical to understand how fabrication errors still do affect the performance of MDLs and numerically evaluate the trade-off between efficiency and initial parameter selection, right at the onset of designing an MDL, i.e., even before putting it into fabrication. Here, we perform a statistical simulation-based study on MDLs (primarily operating in the THz regime) to analyze the impact of various fabrication imperfections (single and multiple) on the final structure as a function of the number of ring height levels. Furthermore, we also evaluate the performance of these same MDLs with the change in the refractive index of the constitutive material. We use focusing efficiency as the evaluation criterion in our numerical analysis; since it is the most fundamental property that can be used to compare and assess the performance of lenses (and MDLs) in general designed for any application with any specific functionality.

1. Introduction

There has been a significant interest in the scientific community in recent times to reduce the thickness and weight of lenses to enable miniature and compact optical systems [1]. Conventional refractive lens, which harness refraction to guide light, fails to satisfy both of this criterion, as it tends to be heavy and bulky owing to its increasing curvature with the increase in numerical aperture [2]. In contrast to this, lenses, which exploit diffraction to guide incident light, have already been shown to be of constant thickness at even larger bending angles in addition to being planar and lightweight. This ability to maintain a constant thickness is simply achieved by decreasing the local period of the diffractive optic [3]. To ensure constructive interference, each incident ray must now locally phase shift to compensate for the variation in its total optical path length to the focal plane. For conventional diffractive lenses, this is achieved by engineering the path traversed by the ray within the diffractive lens itself [3, 4].

In terms of performance, with respect to its refractive counterparts, traditional blazed, or diffractive lenses with almost optimal continuous phase distribution have already been shown to achieve 100% efficiency [5]. Nonetheless, at high numerical apertures, a drop in efficiency occurs due to resonance conditions. In addition to this, conventional diffractive lenses have poor broadband performance due to significant chromatic aberrations. The first problem is mitigated with parametric optimization of constituent elements of the diffractive lens, and the second problem is avoided to some extent through harmonic phase shifts [6] and by using higher orders of diffraction [7]. However, a harmonic diffractive lens (or multi-order diffractive lens) is, in principle, a hybrid refractive-diffractive lens. Such a refractive-diffractive based approach is limited only to a discrete number of operating wavelengths. From such a perspective, multilevel diffractive lenses (MDLs), as shown in **Fig. 1(a)**; *especially broadband MDLs, are fundamentally*

*very different as they can be designed using the same principle of parametric optimization of the constituent elements; but operate across a continuous bandwidth at both low and high numerical apertures with high efficiency.*

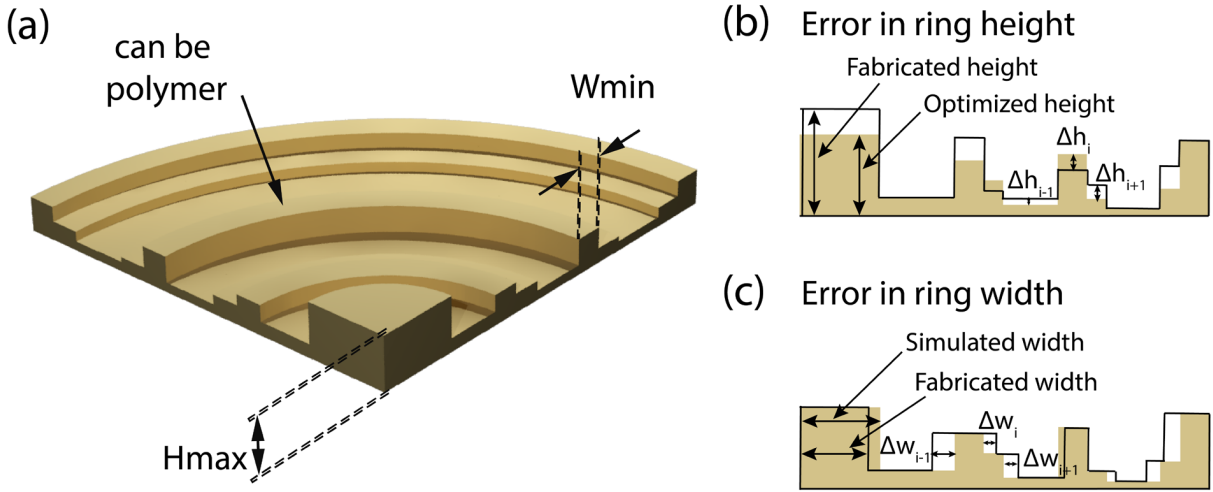

**Fig. 1.** (*a*) Schematic of a THz Multilevel Diffractive Lens (MDL). Characterization of fabrication error in terms of (*b*) error in ring height between the optimized height and the fabricated height and (*c*) error in ring width between the simulated width and the fabricated width of the final MDL structure.

Subwavelength structures have also recently been used to create planar lenses for achromatic focusing and other functionalities [4, 8-10]. These "metalenses" achieve abrupt phase shifts via resonance effects. The metalenses, however, are far more complex to fabricate than MDLs, which is evidenced by the fact that MDLs typically are constituted by lateral features >λ and aspect ratios of 1:3, whereas most metalenses require lateral features <<0.1λ in size and aspect ratios of 3:1 or higher [3]. Nonetheless, diffractive lenses are more than enough to manipulate the scalar properties of light, while metalenses and metasurfaces have an important advantage in terms of controlling the vector properties of light. Therefore, MDLs are implicitly polarization-insensitive due to its scalar nature, whereas metalenses are polarization sensitive. Lastly, MDLs can achieve the same or better optical performance when compared to metalenses. Hence, employing metalenses to model scalar properties of light can be considered "*an overkill*", from all the perspective of theoretical modeling, fabrication as well as performance [3, 11].

These advantages have already been exploited by research groups to design achromatic MDLs via careful parametric optimization of the lens surface topography in the visible [3, 11-22], NIR [23], SWIR, [24] LWIR [25], THz [26, 27] and microwave [28] bands. In fact, we have recently shown the design of a single achromatic MDL with a focal length of 18mm and aperture of ~1mm, operating across a continuous spectrum of wavelengths from 450nm to 15μm [29, 30]. We hypothesize that the achievable bandwidth with MDLs is "unlimited" for practical purposes and is only constrained by the quantum efficiency of the image sensor. Furthermore, we also showcased MDLs with a Field of View (FOV) up to 50° for wide-angle imaging [23] as well as MDLs with a Depth of Focus (DOF) imaging of up to 6 meters in the NIR [31]. Apart from this, MDLs have also been utilized to create broadband holograms enabling multi-plane image projection [32] and in holographic displays for AR/VR applications [33]. Computational imaging with single and multi-aperture MDLs have also been showcased [34-36] along with its potential for applications in photovoltaics [37].

Because of this stupendous progress made using MDLs, a common concern amongst most of the work in this field is the fact that there is still a discrepancy among simulated and experimental efficiency values. This ultimately affects the system performance. We acknowledge the fact that researchers indeed have made efforts to justify these discrepancies by resorting to fabrication errors, as is associated with a non-industrial grade fabrication facility. Yet still, we firmly believe that it would be of immense help to designers to understand how (and in which cases) fabrication errors do affect the performance of MDLs at the initial stage (i.e., even before putting them into fabrication) and numerically evaluate the trade-off between efficiency and initial parameter selection. This will not only help to make a judicious decision while choosing the initial parameters when designing a MDL, keeping in mind the fabrication capability at the corresponding location, but will enable one to predict (at least provide the ballpark estimates of) efficiency even if a certain MDL is designed and tested out.

To facilitate with this process of providing a suitable metric, in this work, we seek to perform a statistical simulation-based study on MDLs (primarily operating in the THz regime) to analyze the impact of various imperfections (single and multiple) on the final structure as a function of the number of ring height levels. **Figures 1(b)** and **1(c)** depict the two major cases, i.e., error in ring height and ring width, which is often commonly encountered in the fabrication of an MDL. Both these cases have been studied individually. Furthermore, an amalgamation of both errors presented simultaneously in the final MDL structure is analyzed, to paint a more practical picture for narrowband and broadband operation. Furthermore, we also evaluate the performance of the same MDLs with variations in the refractive index. An aspect from which error (primarily at the modeling stage) can also occur (e.g. due to variations in density when 3D printing) and propagate during the optimization phase. It is essential to study the variation of performance with respect to material dispersion because most often, it is difficult to ascertain the exact optical constants (refractive index and absorption coefficient) of the material that is going to be used in the manufacturing process. In that case, MDL designers are left to rely upon values of materials already provided in the pre-existing literature while doing the design. In addition to this, many groups or fabrication facilities use modified hybrids of the same standard material, to which other groups do not have access too. A study pertaining to this specific problem is done in [38] pertaining to 3D printable materials used in manufacturing THz diffractive optical elements at a common fabrication facility vs. the author's own laboratory. Finally, even if the design is done with a specific material; to have an idea of how well the same design may perform with a different material with similar properties will help one to gauge the robustness of the designed structure.

The rationale behind the choice of focusing efficiency as the evaluation metric stems from the fact that focusing efficiency is the most basic criterion, which guides the field of lens design for any desired application with any tailored functionality. Other metrics like EDOF, FOV magnification, aberrations can also be included; but these are not often the principal criterion that one looks at when designing an optical element. The choice of performing this study on MDLs primarily operating the THz regime is due to two main reasons [39-42]. First, the ease in modeling larger unit cells leads to the total number of elements in the entire structure to be small and tractable; hence, it is easy to capture the trade-off in efficiency with faster simulations and develop accurate prediction metrics. Therefore, the THz regime is an ideal frequency band to capture the real significance of this work [41, 42]. Second, since the THz band lies in almost the center of the electromagnetic spectrum, hence the metrics developed in this paper can be scaled up or down to both the microwave as well as optical ranges without any loss of generality.

## 2. Results and Discussion

The rotationally symmetric MDLs were designed via a non-linear search method, namely Gradient Descent Assisted Binary Search (GDABS) algorithm. Full description of the design process is already explained in [26, 27]. Therefore, we choose to omit an in-depth discussion of the same here. However, to surmise, each of the designed MDL consists of concentric rings of width equal to a pre-defined value with the ultimate objective of maximizing the focusing efficiency across its bandwidth of operation. Radial symmetry of these structures is exploited to speed up the computation and reduce the optimization time. For our current study, we designed five different MDLs for both narrowband (0.2 THz) as well as broadband (0.1 THz to 0.3 THz) operation. Therefore, in total, ten different MDLs were designed. The ring height distribution and the relevant geometric design parameters are provided in the *supplementary information*.

Speaking of geometric design parameters, the design values were chosen, keeping in mind the fabrication constraints associated with any off-the-shelf hobbyist 3D printer available in the market. This was done to make the study easily comprehensible and understandable for the readers. Each MDL was designed to be 24 mm in diameter. The MDLs designed to operate at a single frequency of 0.2 THz comprised of multilevel concentric rings having a maximum thickness ($h_{max}$) = 1.6 mm whereas the broadband MDLs have a maximum thickness ($h_{max}$) = 3.2 mm. Each ring of the designed MDLs has a width (w) = 0.4 mm. Therefore, the total number of such rings within each MDL = 60. For the purpose of this study, the number of distinct ring height levels were varied from P = [8, 16, 32, 64, 128] which dictated the minimum thickness ($h_{min}$) = [0.2 mm, 0.1 mm, 0.05 mm, 0.025 mm, 0.0125 mm] for narrowband MDLs and ($h_{min}$) = [0.4 mm, 0.2 mm, 0.1 mm, 0.05 mm, 0.025 mm] for broadband MDLs. The focal length was fixed at f = 50 mm, which translates to a numerical aperture (N.A.) of 0.2334. The material chosen for the design of the MDLs was PLA, and its dispersion values (which was inputted into the optimization algorithm) are provided in the *supplementary information.* The reason behind choosing PLA is due to its widespread availability as a common 3D printable material and its negligible loss within the frequency of operation of the designed MDLs. However, other materials could also have been used.

A statistical standard deviation-based simulation error model like in [14, 29, 30] was undertaken to study the impact of fabrication error (ring height and ring width) due to the following reasons: One, the standard deviation is always considered in relation to the mean (or average) since the mean by itself is usually not very useful. Two, the standard deviation is the best measure of variation since it is based on every item in the distribution [43]. Third, the standard deviation is less affected by fluctuations of sampling than most other measures of dispersion. Fourth, the standard deviation is most prominently used in carrying out further statistical studies like computing skewness, correlation, etc. [44]. Therefore, the use of a standard deviation-based error model seems justified. In our case, we varied the ring height ($\Delta h_{err(i)}$) as well as the ring width ($\Delta w_{err(i)}$) of the optimized MDLs first individually and then as an amalgamation of both, and calculated the focusing efficiency at the focal plane using the Fresnel-Kirchoff diffraction integral shown below:

$$U(x',y',\lambda,d) = \frac{e^{ikd}}{i\lambda d} \iint T(x,y,\lambda) \cdot e^{i\frac{k}{2d}[(x-x')^2 + (y-y')^2]} dx dy \tag{1}$$

where d is the focal length, $(x, y)$ are coordinates in the lens plane, $(x', y')$ are coordinates in the focal plane, $and\ T(x, y, \lambda)$ is the pupil function of the MDL. For each number of distinct ring height value P, we recorded ten sets of observation data. From this observation data, we then proceeded to calculate the mean as well as the 90% confidence interval around this mean. We then plotted the same in **Fig. 2** and **Fig. 3**.

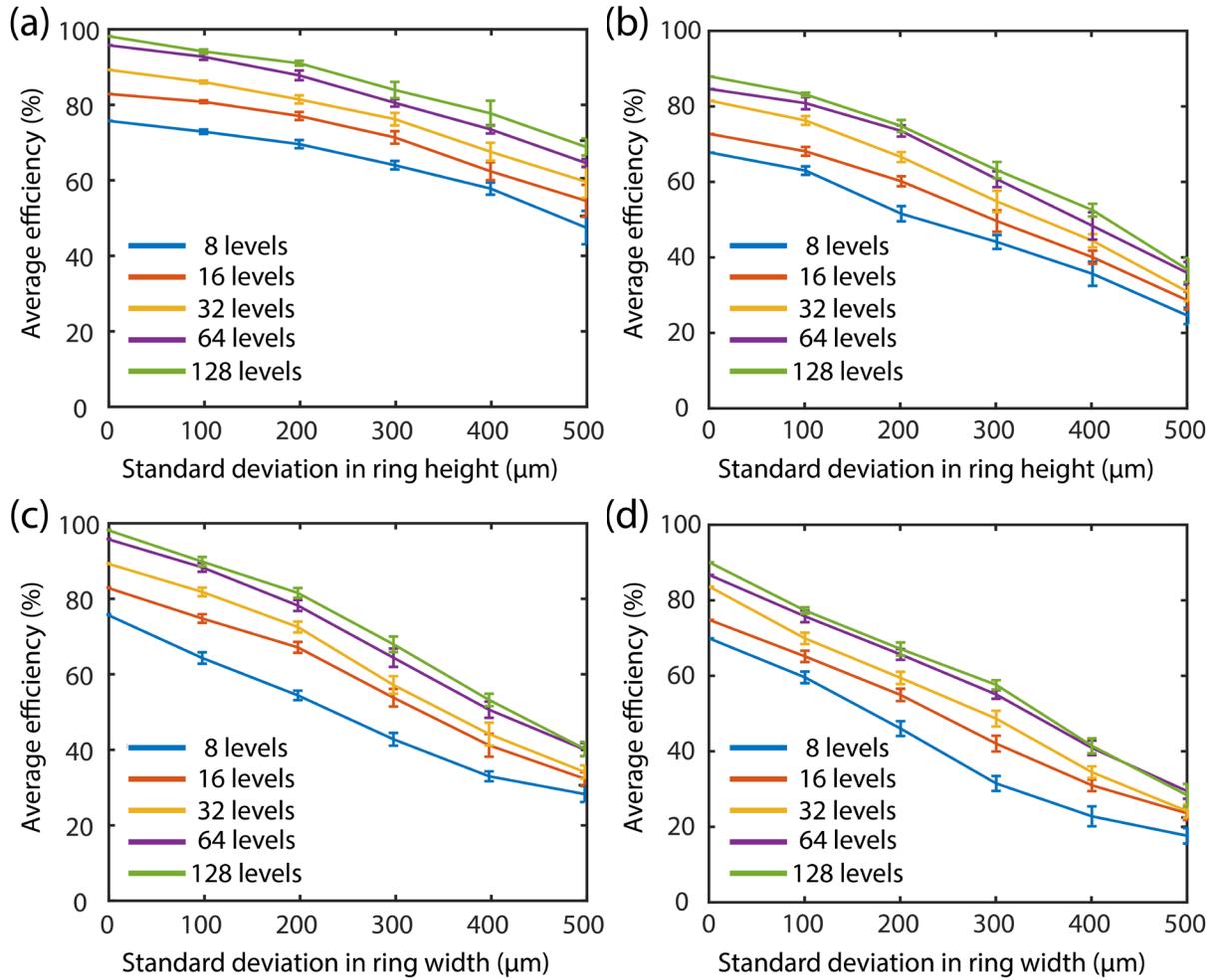

**Fig. 2.** Effect on average focusing efficiency due to a standard deviation-based error in ring height for (*a*) narrowband operation at 0.2 THz and (*b*) broadband operation from 0.1 THz to 0.3 THz. A similar approach is undertaken to characterize the impact due to a standard deviation-based error in ring width under (*c*) narrowband operation at 0.2 THz and (*d*) broadband operation from 0.1 THz to 0.3 THz.

**Fig. 2(a-b)** depicts the impact on average focusing efficiency for the first fabrication error scenario, i.e., standard deviation-based error in ring height for the designed MDLs under narrowband (0.2 THz) and broadband operation (0.1 THz to 0.3 THz) respectively. We computed the focusing

efficiency of the MDLs as the power within a spot of diameter equal to 3 times the FWHM of the spot divided by the total power incident on the lens [3]. Later we averaged the individual focusing efficiencies over the total number of design frequencies. On similar lines, **Fig. 2(c-d)** portrays the impact on average focusing efficiency second fabrication error scenario, i.e., standard deviation-based error in ring width for the designed MDLs under both narrowband (0.2 THz) and broadband operation (0.1 THz to 0.3 THz) correspondingly. We would like to remind the readers that the terms "*average focusing efficiency*" and "*average efficiency*" are the same and will be used interchangeably in this discussion. Moreover, the term "*average efficiency*" for the narrowband MDL is just "*efficiency*" since the number of frequency samples in this case = 1. Moving ahead, we considered the standard deviation in the error of both ring height and width up to 500 μm because the fall in efficiency for all the plots is ~50%, barring only the first case for narrowband MDLs operating at a single frequency.

Some general observations from the plots in **Fig. 2** are as follows: The MDLs designed to operate at only a single frequency are very resilient to errors in both ring height and width as compared to its broadband counterparts. This is easy to understand given the fact that the constraint of focusing a larger number of frequencies at a single point in the observation plane requires a higher degree of control by the diffractive structures, and hence a slight mismatch would significantly degrade its performance. An important corollary to the former statement is the fact that in general the average focusing efficiency of the structures irrespective of whether under narrowband or broadband operation, is higher with the increase in the number of ring height levels. There is a noticeable increase in average efficiency when the number of ring height levels increase from 8 to 64 levels. However, the increase in efficiency gets somewhat saturated for anything over 64 levels. The explanation for this can be attributed to the fact that at 64 levels, the surface topography of the MDL is almost continuous, i.e., resembles a blazed grating structure and hence, any increase in the number of height levels beyond 64 levels does not result in a steep rise in efficiency. This effect is like what is observed in conventional multilevel DOEs [5].

The third important observation is that the effect of errors in the width of the designed MDLs is in general greater than the effect of error in height under both narrowband (0.2 THz) as well as broadband operation (0.1 THz to 0.3 THz). The reason for this behaviour, we believe is because only the ring height profile is optimized for the MDL structures with $\Delta h$ = [0.2 mm, 0.1 mm, 0.05 mm, 0.025 mm, 0.0125 mm] for narrowband MDLs and $\Delta h$ = [0.4 mm, 0.2 mm, 0.1 mm, 0.05 mm, 0.025 mm] for broadband MDLs with P = [8,16,32,64,128]; whereas the width of each ring is kept constant at 0.4 mm during the entire optimization. Even if one assumes that the optimized MDL might converge to a solution where the height of two adjacent rings have the same value such that both the rings can now be considered as a single ring of 0.8 mm thickness, the differential width ($\Delta w$) still has a smaller degree of freedom in contrast to its height. Finally, to quantify the results, for a 20% reduction in focusing efficiency, the narrowband MDLs require a ~350 µm error in ring height as compared to a ring width error of ~250 µm. For the broadband MDLs, the values are ~250 µm and ~200 µm in ring height and width, respectively. The "20% decrease in focusing efficiency" metric has been adopted from [3], where it is seen that the change in the PSF plots is prominent for this value of decease in focusing efficiency.

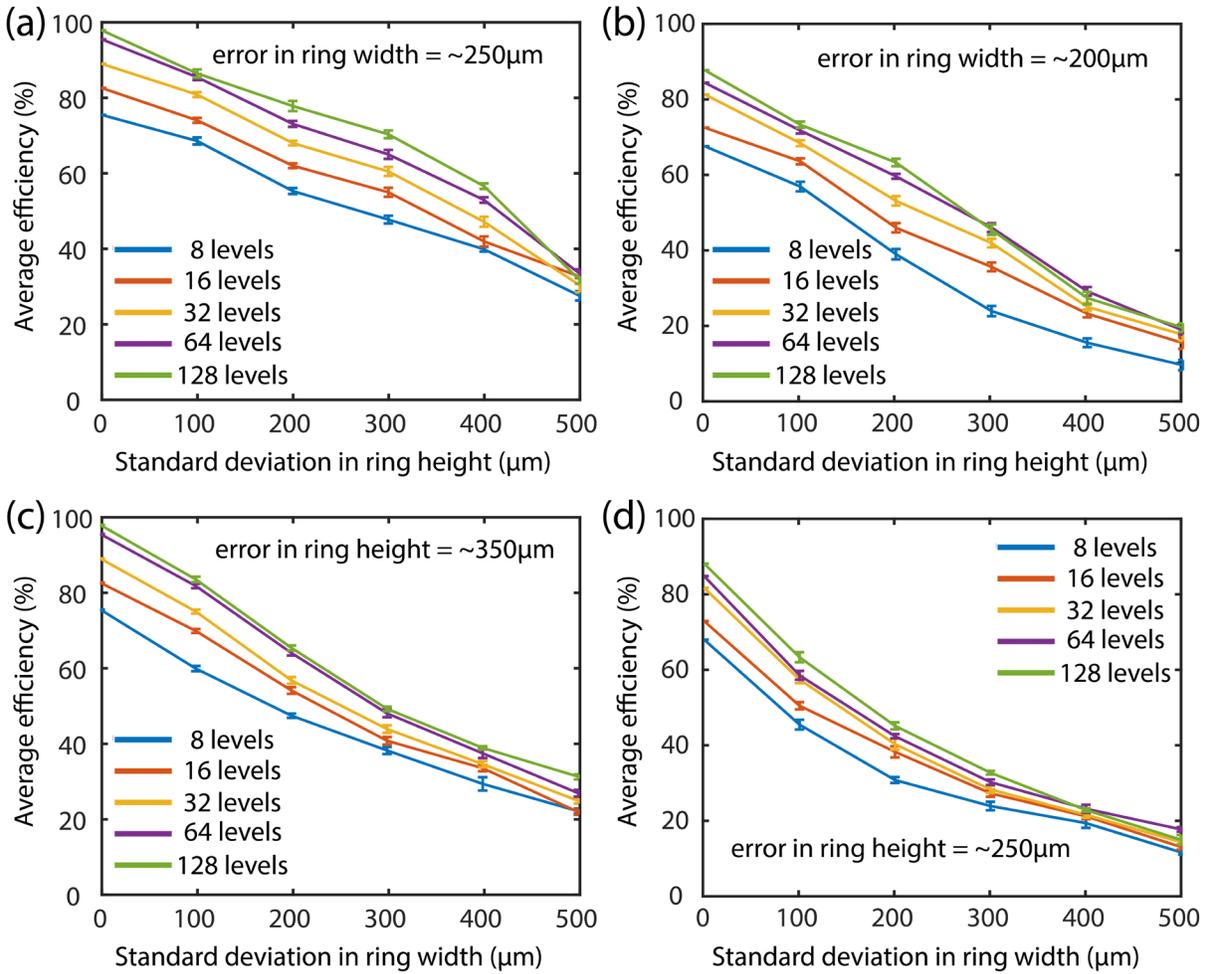

**Fig. 3.** Impact on average efficiency due to a standard deviation-based error in ring height for a fixed error in (*a*) width = ~250 μm under narrowband operation at 0.2 THz and (*b*) width = ~200 μm under broadband operation from 0.1 THz to 0.3 THz. Consequently, the impact due to a standard deviation-based error in ring width for a fixed error in (*c*) height = ~350 μm under narrowband operation at 0.2 THz and (*d*) height = ~250 μm under broadband operation from 0.1 THz to 0.3 THz.

**Fig. 3** depicts a more complicated yet practical scenario of fabrication errors occurring in MDLs. To be specific, **Fig. 3(a-b)** now shows the effect on average efficiency of the designed MDLs by varying the ring heights as per the standard deviation-based error model for a fixed error in width under both narrow and broadband case. An almost same analogy is adopted for the plots in **Fig. 3(c-d)** where now instead, the ring widths are varied for a fixed error in ring height under both narrowband and broadband conditions. Similar observations to those in **Fig. 2** are witnessed even in the plots of **Fig. 3** with two key differences. One, the fall in average focusing efficiency is sharper for both the narrowband as well as the broadband MDLs. We took the fixed error values for ring width and height from the observations in **Fig. 2**. Two, the 90% confidence intervals in the plots of **Fig. 3** overlap in the region of standard deviation based errors of "400 μm - 500 μm" which intuitively tells us the design solutions start to become erratic and unstable (i.e., an MDL with P =128 might have a lower efficiency than an MDL with P = 32 under broadband operation). This is even seen in the plots of **Fig. 2**, but the effect is not as prominent, and the overlap is only marginally observed for **Fig. 2(d)**. Lastly, to quantify the results for the plots in **Fig. 3**, the narrowband MDLs require a ~200 μm error in ring height as compared to a ring width error of ~150 μm for a similar drop in 20% of focusing efficiency. For the broadband MDLs, the values are ~150 μm and ~100 μm in ring height and width, respectively. Therefore, we see that the combined impact of error in both ring height and ring width is even more significant than the error contributed by only ring height or width. This was expected and gives an estimate of what initial parameters to choose from while designing an MDL, taking into consideration the number of errors that can be introduced by the 3D printer that will be employed to fabricate them.

Finally, we study the impact of change in the refractive index of the constitutive material on the average focusing efficiency of the MDL structures, again, at both a single frequency of 0.2 THz as well as a broadband regime from 0.1 THz to 0.3 THz. The respective plots for both these cases are depicted in **Fig. 4(a)** and **Fig. 4(b)**, respectively. A point of note here is that the absorption coefficient is very negligible in this frequency range, i.e., 0.1 THz to 0.3 THz, as is evidenced from the plot in **Fig. S2** of the ***supplementary information***. Hence, when the refractive index values were changed during this study, the absorption coefficient was updated accordingly. This was done to keep the error model understandable and straightforward. From **Fig. 4(b)** it is observed that the fall in the average focusing efficiency is steeper for the broadband MDLs in contrast to the narrowband MDLs in **Fig. 4(b)**. The narrowband MDLs are also more resilient to the change in refractive index too. This, again, can be attributed to the "*potential challenge*" in handling more design constraints with accuracy. Another important observation that is critical here is the fact that for the change in refractive index ($n_\lambda \pm 1$ to $n_\lambda \pm 2$), the MDL solutions for both narrowband and broadband become equally unstable. This tells us that irrespective of the bandwidth or the number of ring height levels, the optimized structures will just not work if the dispersion values are way off than what was used during the design phase.

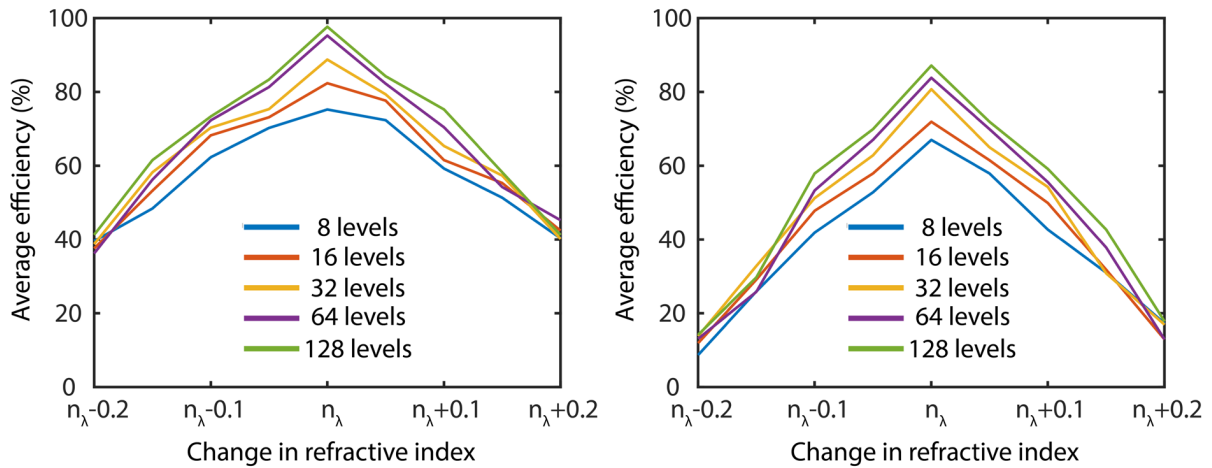

**Fig. 4.** Effect on average focusing efficiency of the MDL structures with the change in the refractive index under (*a*) narrowband operation at 0.2 THz and (*b*) broadband operation from 0.1 THz to 0.3 THz, respectively.

## 3. Conclusion

In conclusion, we have performed a statistical simulation-based study on MDLs designed to operate in the THz regime to analyze the impact of various fabrication imperfections in the final structure as a function of the number of ring height levels. In addition to this, we also analyzed the performance of these same MDLs with the change in the refractive index. We firmly believe that this work provides important information to MDL designers and researchers alike to compare and assess the performance of MDLs as well as offer fundamental insight into designing highly efficient and robust MDLs with given fabrication constraint.

**Acknowledgments**

This work was supported by the NSF awards: ECCS # 1936729 and MRI #1828480

# Supplementary Information
# Role of Fabrication Errors and Refractive Index on Multilevel Diffractive Lens Performance

Sourangsu Banerji[1], Jacqueline Cooke[1] and Berardi Sensale-Rodriguez[1, *]

[1] Department of Electrical and Computer Engineering, The University of Utah, Salt Lake City, UT 84112, USA

* E-mail: berardi.sensale@utah.edu

1. **Refractive index and absorption co-efficient of PLA**

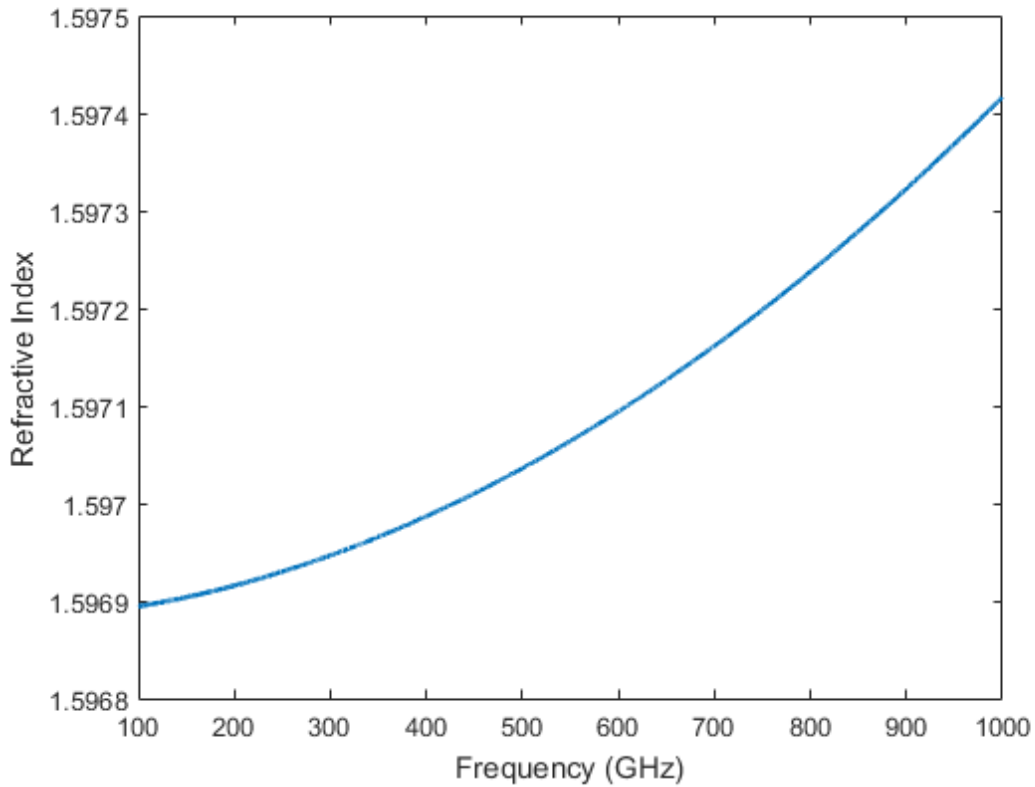

**Fig. 1.** Refractive Index of PLA polymer

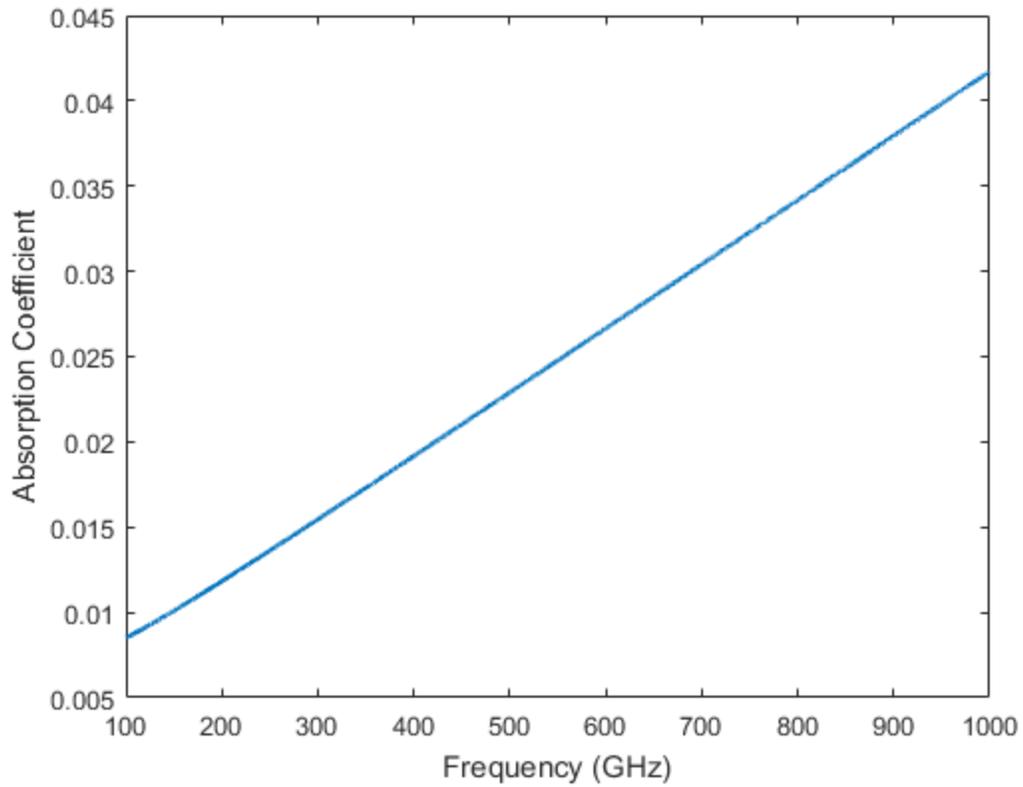

**Fig. 2.** Absorption co-efficient of PLA polymer

## 2. MDL designs

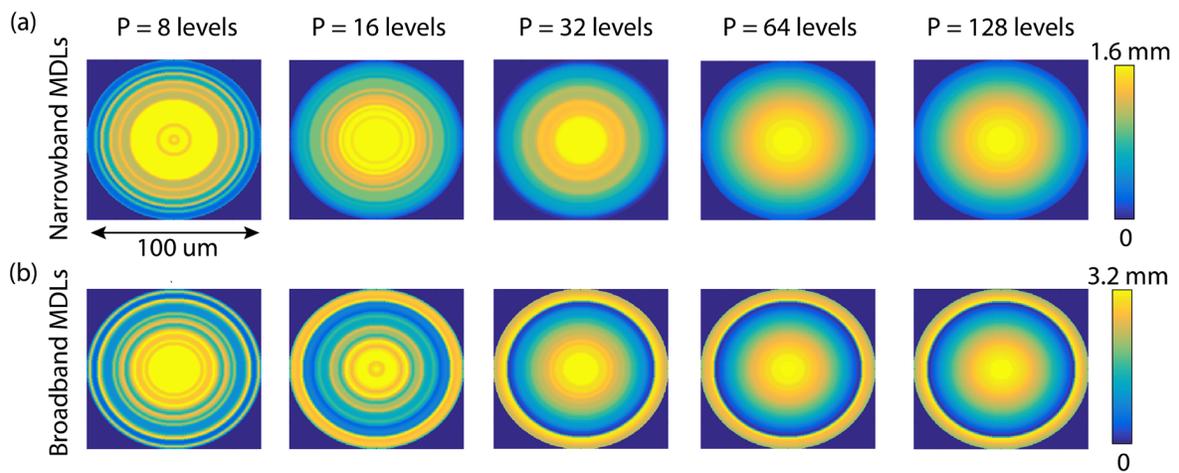

**Fig. 3.** MDL designs for P = [8, 16, 32, 64, 128] for **(a)** narrowband and **(b)** broadband